%
\magnification=1200\overfullrule=0pt\baselineskip=15pt
\hoffset=1cm
\voffset=1.5cm
\vsize=22truecm \hsize=15truecm \overfullrule=0pt\pageno=0

\font\titlefont=cmbx10 scaled \magstep1
\font\sectnfont=cmbx8 scaled \magstep1
\def\mname{\ifcase\month\or January \or February \or March \or April
           \or May \or June \or July \or August \or September
           \or October \or November \or December \fi}
\def\date{\hbox{\strut\mname \number\year}}
\def\binum{\hbox{BI-TP 94/30\strut}}
\def\banner{\hfill\hbox{\vbox{\offinterlineskip\binum\date}}\relax}
\def\manner{\hbox{\vbox{\offinterlineskip}}
               \hfill\relax}
\footline={\ifnum\pageno=0\manner\else\hfil\number\pageno\hfil\fi}
%

\def\PRL{{ Phys.\ Rev.\ Lett.\ }}
\def\PL{{ Phys.\ Lett.\ }}
\def\NP{{ Nucl.\ Phys.\ }}
%
%
%
\newcount\FIGURENUMBER\FIGURENUMBER=0
\def\fig#1{\expandafter\ifx\csname FG#1\endcsname\relax
               \global\advance\FIGURENUMBER by 1
               \expandafter\xdef\csname FG#1\endcsname
                              {\the\FIGURENUMBER}\fi
           Fig.~\csname FG#1\endcsname\relax}
\def\figs#1#2{\expandafter\ifx\csname FG#1\endcsname\relax
               \global\advance\FIGURENUMBER by 1
               \expandafter\xdef\csname FG#1\endcsname
                              {\the\FIGURENUMBER}\fi
              \expandafter\ifx\csname FG#2\endcsname\relax
               \global\advance\FIGURENUMBER by 1
               \expandafter\xdef\csname FG#2\endcsname
                              {\the\FIGURENUMBER}\fi
           Figs.~\csname FG#1\endcsname and
                       \csname FG#2\endcsname\relax}
\def\Fig#1{\expandafter\ifx\csname FG#1\endcsname\relax
               \global\advance\FIGURENUMBER by 1
               \expandafter\xdef\csname FG#1\endcsname
                              {\the\FIGURENUMBER}\fi
           Figure \csname FG#1\endcsname\relax}
%
\newcount\TABLENUMBER\TABLENUMBER=0
\def\table#1{\expandafter\ifx\csname TB#1\endcsname\relax
               \global\advance\TABLENUMBER by 1
               \expandafter\xdef\csname TB#1\endcsname
                               {\the\TABLENUMBER}\fi
             Table \csname TB#1\endcsname\relax}
\def\Table#1{\expandafter\ifx\csname TB#1\endcsname\relax
               \global\advance\TABLENUMBER by 1
               \expandafter\xdef\csname TB#1\endcsname
                               {\the\TABLENUMBER}\fi
             TABLE \csname TB#1\endcsname\relax}
%
\newcount\EQUATIONNUMBER\EQUATIONNUMBER=0
\def\EQNO#1{\expandafter\ifx\csname EQ#1\endcsname\relax
               \global\advance\EQUATIONNUMBER by 1
               \expandafter\xdef\csname EQ#1\endcsname
                            {\the\EQUATIONNUMBER}\fi
            \eqno(\csname EQ#1\endcsname)\relax}
\def\EQNAME#1{\expandafter\ifx\csname EQ#1\endcsname\relax
               \global\advance\EQUATIONNUMBER by 1
               \expandafter\xdef\csname EQ#1\endcsname
                            {\the\EQUATIONNUMBER}\fi
            (\csname EQ#1\endcsname)\relax}
\def\eq#1{\expandafter\ifx\csname EQ#1\endcsname\relax
               \global\advance\EQUATIONNUMBER by 1
               \expandafter\xdef\csname EQ#1\endcsname
                           {\the\EQUATIONNUMBER}\fi
          Eq.~(\csname EQ#1\endcsname)\relax}
\def\eqand#1#2{\expandafter\ifx\csname EQ#1\endcsname\relax
               \global\advance\EQUATIONNUMBER by 1
               \expandafter\xdef\csname EQ#1\endcsname
                        {\the\EQUATIONNUMBER}\fi
          \expandafter\ifx\csname EQ#2\endcsname\relax
               \global\advance\EQUATIONNUMBER by 1
               \expandafter\xdef\csname EQ#2\endcsname
                      {\the\EQUATIONNUMBER}\fi
         Eqs.~(\csname EQ#1\endcsname{})
         and (\csname EQ#2\endcsname)\relax}
\newcount\REFERENCENUMBER\REFERENCENUMBER=0
\def\REF#1{\expandafter\ifx\csname RF#1\endcsname\relax
               \global\advance\REFERENCENUMBER by 1
               \expandafter\xdef\csname RF#1\endcsname
                         {\the\REFERENCENUMBER}\fi}
\def\reftag#1{\expandafter\ifx\csname RF#1\endcsname\relax
               \global\advance\REFERENCENUMBER by 1
               \expandafter\xdef\csname RF#1\endcsname
                      {\the\REFERENCENUMBER}\fi
             \csname RF#1\endcsname\relax}
\def\ref#1{\expandafter\ifx\csname RF#1\endcsname\relax
               \global\advance\REFERENCENUMBER by 1
               \expandafter\xdef\csname RF#1\endcsname
                      {\the\REFERENCENUMBER}\fi
             [\csname RF#1\endcsname]\relax}
\def\refto#1#2{\expandafter\ifx\csname RF#1\endcsname\relax
               \global\advance\REFERENCENUMBER by 1
               \expandafter\xdef\csname RF#1\endcsname
                      {\the\REFERENCENUMBER}\fi
           \expandafter\ifx\csname RF#2\endcsname\relax
               \global\advance\REFERENCENUMBER by 1
               \expandafter\xdef\csname RF#2\endcsname
                      {\the\REFERENCENUMBER}\fi
             [\csname RF#1\endcsname--\csname RF#2\endcsname]\relax}
\def\refs#1#2{\expandafter\ifx\csname RF#1\endcsname\relax
               \global\advance\REFERENCENUMBER by 1
               \expandafter\xdef\csname RF#1\endcsname
                      {\the\REFERENCENUMBER}\fi
           \expandafter\ifx\csname RF#2\endcsname\relax
               \global\advance\REFERENCENUMBER by 1
               \expandafter\xdef\csname RF#2\endcsname
                      {\the\REFERENCENUMBER}\fi
            [\csname RF#1\endcsname,\csname RF#2\endcsname]\relax}
\def\refss#1#2#3{\expandafter\ifx\csname RF#1\endcsname\relax
               \global\advance\REFERENCENUMBER by 1
               \expandafter\xdef\csname RF#1\endcsname
                      {\the\REFERENCENUMBER}\fi
           \expandafter\ifx\csname RF#2\endcsname\relax
               \global\advance\REFERENCENUMBER by 1
               \expandafter\xdef\csname RF#2\endcsname
                      {\the\REFERENCENUMBER}\fi
           \expandafter\ifx\csname RF#3\endcsname\relax
               \global\advance\REFERENCENUMBER by 1
               \expandafter\xdef\csname RF#3\endcsname
                      {\the\REFERENCENUMBER}\fi
[\csname RF#1\endcsname,\csname RF#2\endcsname,\csname
RF#3\endcsname]\relax}
\def\refand#1#2{\expandafter\ifx\csname RF#1\endcsname\relax
               \global\advance\REFERENCENUMBER by 1
               \expandafter\xdef\csname RF#1\endcsname
                      {\the\REFERENCENUMBER}\fi
           \expandafter\ifx\csname RF#2\endcsname\relax
               \global\advance\REFERENCENUMBER by 1
               \expandafter\xdef\csname RF#2\endcsname
                      {\the\REFERENCENUMBER}\fi
            [\csname RF#1\endcsname,\csname RF#2\endcsname]\relax}
\def\Ref#1{\expandafter\ifx\csname RF#1\endcsname\relax
               \global\advance\REFERENCENUMBER by 1
               \expandafter\xdef\csname RF#1\endcsname
                      {\the\REFERENCENUMBER}\fi
             [\csname RF#1\endcsname]\relax}
\def\Refto#1#2{\expandafter\ifx\csname RF#1\endcsname\relax
               \global\advance\REFERENCENUMBER by 1
               \expandafter\xdef\csname RF#1\endcsname
                      {\the\REFERENCENUMBER}\fi
           \expandafter\ifx\csname RF#2\endcsname\relax
               \global\advance\REFERENCENUMBER by 1
               \expandafter\xdef\csname RF#2\endcsname
                      {\the\REFERENCENUMBER}\fi
            [\csname RF#1\endcsname--\csname RF#2\endcsname]\relax}
\def\Refand#1#2{\expandafter\ifx\csname RF#1\endcsname\relax
               \global\advance\REFERENCENUMBER by 1
               \expandafter\xdef\csname RF#1\endcsname
                      {\the\REFERENCENUMBER}\fi
           \expandafter\ifx\csname RF#2\endcsname\relax
               \global\advance\REFERENCENUMBER by 1
               \expandafter\xdef\csname RF#2\endcsname
                      {\the\REFERENCENUMBER}\fi
        [\csname RF#1\endcsname,\csname RF#2\endcsname]\relax}
%
\newcount\SECTIONNUMBER\SECTIONNUMBER=0
\def\section#1{\global\advance\SECTIONNUMBER by 1
      \bigskip\goodbreak\line{{\sectnfont \the\SECTIONNUMBER.\ #1}\hfil}
      \smallskip}
%
\def\DT{\Delta\tau}\def\dt{\ifmmode\DT\else$\DT$\fi}
\def\BETAC{\beta_c}\def\betac{\ifmmode\BETAC\else$\BETAC$\fi}
\def\LARGE{8\times16^3}\def\large{\ifmmode\LARGE\else$\LARGE$\fi}
\def\SMALL{8\times12^3}\def\small{\ifmmode\SMALL\else$\SMALL$\fi}
\def\MQ{ma}\def\mq{\ifmmode\MQ\else$\MQ$\fi}
\def\REL{Re(L)}\def\rel{\ifmmode\REL\else$\REL$\fi}
\def\ABL{Abs(L)}\def\absl{\ifmmode\ABL\else$\ABL$\fi}
\def\PPBAR{\overline\chi\chi}\def\ppbar{\ifmmode\PPBAR\else$\PPBAR$\fi}
\def\ord{\ifmmode\langle\REL\rangle\else$\langle\REL\rangle$\fi}
\def\cord{\ifmmode\langle\PPBAR\rangle\else$\langle\PPBAR\rangle$\fi}
\def\pprgi{\ifmmode\langle\PPBAR\rangle_{RGI}\else$\langle\PPBAR
           \rangle_{RGI}$\fi}
\def\entropy{\ifmmode{s\over T^3}
             \else$s/T^3$\fi}
\def\NT{N_\tau}\def\nt{\ifmmode\NT\else$\NT$\fi}
\def\NS{N_\sigma}\def\ns{\ifmmode\NS\else$\NS$\fi}
\def\GT{g_\tau}\def\gt{\ifmmode\GT\else$\GT$\fi}
\def\GS{g_\sigma}\def\gs{\ifmmode\GS\else$\GS$\fi}
\def\CT{c'_\tau}\def\ct{\ifmmode\CT\else$\CT$\fi}
\def\CS{c'_\sigma}\def\cs{\ifmmode\CS\else$\CS$\fi}
\def\NF{n_\f}\def\nf{\ifmmode\NF\else$\NF$\fi}
\def\TC{T_c}\def\tc{\ifmmode\TC\else$\TC$\fi}
\def\nop{non-perturbative}

\def\lsim{\raise0.3ex\hbox{$<$\kern-0.75em\raise-1.1ex\hbox{$\sim$}}}
\def\gsim{\raise0.3ex\hbox{$>$\kern-0.75em\raise-1.1ex\hbox{$\sim$}}}
%

\def\f{{\scriptscriptstyle f}}
%

\def\ie{{i.~e.\/}}
\def\eg{{e.~g.\/}}
%
%
%
\banner\bigskip\begingroup\titlefont\obeylines
\vskip 20pt
\centerline{SCALING PROPERTIES OF THE}
\vskip 0.3 truecm
\centerline{ENERGY DENSITY IN}
\vskip 0.3 truecm
\centerline{SU(2) LATTICE GAUGE THEORY$^*$}
\vskip 10pt
\endgroup\bigskip
\bigskip
\centerline{J.~Engels$^{1}$, F.~Karsch$^{1}$, and K.~Redlich$^{1,2}$}
\bigskip
\footnote{}{~\par
\noindent1~~Fakult\"at f\"ur Physik, Universit\"at Bielefeld, D-33615
Bielefeld, Germany~\par
\noindent2~~Institute for Theoretical Physics, University of Wroc\l aw,
PL-50205 Wroc\l aw, Poland}
\footnote{}{~\par\noindent$^*$Work supported by the
Deutsche Forschungsgemeinschaft under research grant Pe 340/3-2
and the Bundesministerium f\"ur Forschung und Technologie (BMFT)
}
%
\bigskip\bigskip\bigskip
\centerline{{\bf ABSTRACT}}\medskip
 The lattice data for the energy density of $SU(2)$ gauge theory
are calculated with \nop~derivatives of the coupling
constants. These derivatives are obtained from two sources :
i) a parametrization of the \nop~beta function in accord with
the measured critical temperature and $\Delta\beta-$values and
ii) a \nop~calculation of the presssure. We then perform a detailed
finite size scaling analysis of the energy density near $T_c$.
It is shown that at the critical temperature the energy density is
scaling as a function of $VT^3$
with the corresponding $3d$ Ising model critical exponents.
The value of $\epsilon(T_c)/T^4_c$ in the continuum limit is
estimated to be 0.256(23). In the high temperature regime the energy
density is approaching its weak coupling limit from below, at
$T/T_c \approx 2$ it has reached only about $70\%$ of the limit.
\vfill\eject

\par\noindent {\bf 1. Introduction}
\bigskip
The energy density for finite temperature $SU(N)$ gauge systems
is essential for the understanding and the investigation of the
deconfinement transition and the plasma phase at high temperatures.
In particular, a reliable estimate of the energy density $\epsilon$
at the critical temperature $T_c$ is needed for the design of heavy
ion collision experiments aiming at the production of the quark-gluon
plasma. For theory \eg, the high temperature behaviour of $\epsilon$
is of interest for the test of and the connection to finite temperature
perturbation theory. The determination of the correct non-perturbative
energy density is therefore an important goal of Monte Carlo
simulations.
\medskip
 The lattice calculations of $\epsilon$ performed so far, though they
are showing the same general behaviour for $\epsilon$ on different
size lattices, are not yet scaling, \ie~they do not lead to a unique
function $\epsilon(T)$.
\medskip
One of the sources of this disagreement has been addressed in
\ref{pressure} and is due to the use of weak coupling approximations
to the derivatives with respect to the lattice anisotropy
$\xi=a/a_{\tau}$ ( $a$ and $a_{\tau}$ are the lattice spacings in
space and time directions ) of the space-like ($g_{\sigma}$) and
time-like ($g_{\tau}$) coupling constants. The derivatives appear in
the lattice expressions for the pressure $P$ and the energy density
\medskip
$$
 {\epsilon + P \over T^4} =
 8N\nt^4g^{-2}\biggl[1-{g^2 \over{2}}
\biggl({{\partial}\gs^{-2} \over{\partial}\xi}
- {{\partial} \gt^{-2} \over {\partial} \xi}\biggr)\biggr]
  (P_{\sigma}-P_{\tau})~;
\EQNO{ep}
$$
$$
 \Delta \equiv{\epsilon - 3P \over T^4} =
 12N\nt^4\biggl({{\partial}\gs^{-2} \over{\partial}\xi}
+ {{\partial} \gt^{-2} \over {\partial} \xi}\biggr)
  \biggl[ 2P_0-(P_{\sigma}+P_{\tau}) \biggr]~.
\EQNO{delta}
$$
\medskip
Here $P_{\sigma,\tau}$ are the space and time plaquettes,
$P_{\sigma,\tau} =
{1 \over N} \langle {\rm Tr} (1- U_1U_2U_3^{\dagger}U_4^{\dagger})
\rangle$, on lattices of size $\ns^3\times\nt$. The plaquette
$P_0$ is measured on an $\ns^4$-lattice
and serves to normalize the interaction measure
$\Delta$ at $T = 0$. The derivatives have to be
taken at $\xi = 1$. Their sum is related to the $\beta$-function
through
$$
\beta_f = -a { dg \over da} = -g^3
\biggl({{\partial}\gs^{-2} \over{\partial}\xi}
+ {{\partial} \gt^{-2} \over {\partial} \xi}\biggr)_{\xi=1}~~.
\EQNO{sumrule}
$$
\medskip
Once the $\beta-$function and the pressure are known non-perturbatively,
the energy density can be calculated with the appropriate \nop~
derivatives. This has been done in \ref{pressure} for $SU(3)$ and
$\ns^3\times4$ lattices. In $SU(2)$, where much more data on lattices
with various $\ns$ and $\nt$ exist, this correction is still lacking.
In the next two sections we shall therefore derive first a
non-perturbative $\beta-$function for $SU(2)$ and then the
\nop~derivatives separately.
\medskip
The other major effect on the energy density is coming from the
finite size dependence on the spatial ($\ns$) and temporal ($\nt$)
extents of the lattice. Here we have to distinguish two regions
of temperature or coupling constants. In the neighbourhood of the
deconfinement transition we expect finite size scaling governed by
the critical exponents of the transition and the increase of the
correlation length.
In the weak coupling or high temperature regime we expect a different
behaviour of the energy density, which is dominated by the weak
coupling expansion of $\epsilon$ and its \ns- and (stronger)
\nt-dependence. We discuss the corresponding finite size equations
in section 4.
\medskip
We shall compare the available $SU(2)$ data for the energy density
after correction with non-perturbative coupling derivatives to the
expectations from finite size scaling theory as described above.
The results  are presented and discussed in section 5.

 \bigskip
\par\noindent
{\bf 2. Beta function and scaling in $SU(2)$ lattice gauge theory}
 \bigskip
 \par
 It is well known, that physical quantities such as the critical
 temperature or the string tension are violating asymptotic scaling
 in the range of coupling constants considered so far in lattice
 calculations. Nevertheless the scaling of dimensionless ratios
 of observables works remakably well \ref{asymp} suggesting that
 deviations from asymptotic scaling may be described by a common
 \nop~ $\beta$-function. To parametrize the corresponding
 \nop~ dependence of the lattice spacing $a$ on $g^2$ we make the
 following ansatz
$$
a\Lambda_L = R(g^2)\cdot\lambda (g^2)~,
\EQNO{ansatz}
$$
where
$$
R(g^2) =
\exp[
-{{b_1}\over {2b_0^2}}\ln (b_0 g^2) -{1\over {2b_0 g^2}}
]
\EQNO{renorm}
$$
is resulting from the integration of the two-loop weak coupling
expansion of the $\beta$-function

$$
-\beta_f = b_0 g^3+ b_1 g^5 + ... ~~.
\EQNO{expans}
$$
\medskip
Here
 $$
 b_0={{11N}\over {48\pi^2}} ~~~,~~~
 b_1={{34}\over 3}({{N}\over {16\pi^2}})^2
\EQNO{bees}
 $$
are scheme independent.
The inclusion of terms of order $O(g^7)$ in the expansion of the
$\beta$-function is then equivalent to a deviation of $\lambda$
from unity.
\medskip
The relation of the correction factor $\lambda$ to the sum of \nop~
derivatives is found by rewriting \eq{sumrule} as
$$
 { dg^{-2} \over d\ln a} = -2
\biggl({{\partial}\gs^{-2} \over{\partial}\xi}
+ {{\partial} \gt^{-2} \over {\partial} \xi}\biggr)_{\xi=1}~~,
\EQNO{sum}
$$
and using
$$
 { d\ln a \over dg^{-2} } = -{1 \over 2b_0} +
 {b_1 \over 2b_0^2}g^2 + { d\ln \lambda \over dg^{-2}}~.
\EQNO{lnl}
$$
It is then convenient to write $\lambda$ for small $g^2$
in the following asymptotic form
$$
\lambda(g^2)=\exp[
{1\over {2b_0^2}} ( c_1g^2 + c_2g^4 + c_3g^6 + ...) ]~,
\EQNO{lambda}
$$
which implies $\lambda(0)=1$.
\medskip
The function $\lambda$ may be determined from any measured quantity
of non-zero dimension. However only for small enough values of $g^2$
one may hope to find a unique function from different observables.
\medskip
 In $SU(2)$ lattice gauge theory we shall exploit the measurements
of the critical coupling constant $g^2_c$ for $\nt=2,3,4,5,6,8,16$
\refss{asymp}{Engc}{Engd} to derive $\lambda$.
 From the $g^2_c$-values
we find the critical temperature $T_c$. On a lattice with $\nt$
points in the temporal direction the temperature is
$$
T = {1 \over {\nt a}}~~.
\EQNO{temp}
$$
Therefore we have

$$
\lambda(g^2_c) = {1 \over {\nt R(g^2_c) \cdot T_c/\Lambda_L}}~.
\EQNO{lambda}
$$
Demanding independence of $g^2$ for $T_c/\Lambda_L$
allows us then to determine the function $\lambda(g^2)$
in the presently studied coupling regime.
 In Fig. 1 we show the critical temperature
$1/(\nt R)$ as calculated from the two-loop formula,
\eq{renorm}, \ie~by assuming $\lambda =1$,
 at the measured critical coupling for various $\nt$.
\par
It has been noted \refss{asymp}{Akemi}{Kh} that the deviations from
asymptotic scaling, which are obvious from Fig. 1 can,
to a large extent, be accounted
for by the introduction of an effective coupling constant.
\par
For our present analysis of bulk thermodynamics it is, however,
more convenient to achieve a parametrization of the scaling
violations in terms of the bare coupling constant. This has the
advantage of giving us directly the $\beta-$function, which we need
anyhow for the evaluation  of the energy density and the
pressure.
\par
We thus proceed in the following way. First we use only those points,
which belong to the weak coupling regime ($4/g^2 > 2.23$,
\ie~$\nt\geq4$) to make a fit to $1/(\nt R)$
with the asymptotic form, \eq{lambda}, of $\lambda$. We have
tried various possibilities for the exponent of $\lambda$ in
\eq{lambda}. A rather good fit is obtained with the simple
parametrization
$$
c_1 \equiv c_2 \equiv c_{n>3} \equiv 0~,
\EQNO{fit1}
$$
so that
$$
\lambda_{as} = \exp \left[ {{c_3g^6} \over {2b_0^2}} \right] ~.
\EQNO{las}
$$
The best result is given by $c_3 = 5.529(63) \cdot 10^{-4}$,
which leads to
$$
T_c/\Lambda_L =21.45(14)~.
\EQNO{fit2}
$$
The last number is compatible with a corresponding extrapolation
to $a=0$ in ref.\ref{asymp}.
\par
The simple asymptotic fit \eq{las} deviates in the neighbourhood
of the crossover point from the measured two-loop
critical temperature
values, as can be seen in Fig. 1. In a second step we have
therefore essentially fitted the measured points with a Spline
interpolation, which continuously extends into the region, where
\eq{las} is valid. The solid line in Fig. 1 shows the final
interpolation and the ratio of the measured
$1/(\nt R(g^2_c))$ over the fitted $\lambda(g^2_c)$.
The ratio should be a constant and equal to $T_c/\Lambda_L$.
In \table{nonpert} we give $\lambda$ and the derivative
from \eq{sum}, which up to a factor (-2) equals to the sum of
derivatives. The latter is shown in Fig. 2.
\medskip

$$\vbox{\offinterlineskip
\halign{
\strut\vrule     \hfil $#$ \hfil  &
      \vrule # & \hfil $#$ \hfil  &
      \vrule # & \hfil $#$ \hfil  &
      \vrule # & \hfil $#$ \hfil  &
      \vrule # & \hfil $#$ \hfil
      \vrule \cr
\noalign{\hrule}
  ~4/g^2~
&&~\lambda~
&&~{adg^{-2}/da}~
&&~{dg_{\sigma}^{-2}/d\xi}~
&&~{dg_{\tau}^{-2}/d\xi}~\cr
\noalign{\hrule}
%
%
~~~2.15~~~&&~~~1.913188~~~&&~~~-.137878~~~&&~~~.571969~~~&&~~~-.503031~~~\cr
~~~2.20~~~&&~~~1.964628~~~&&~~~-.117386~~~&&~~~.377374~~~&&~~~-.318682~~~\cr
~~~2.25~~~&&~~~1.983554~~~&&~~~-.099932~~~&&~~~.283626~~~&&~~~-.233660~~~\cr
~~~2.26~~~&&~~~1.982803~~~&&~~~-.096746~~~&&~~~.265652~~~&&~~~-.217279~~~\cr
~~~2.27~~~&&~~~1.980406~~~&&~~~-.093647~~~&&~~~.246364~~~&&~~~-.199541~~~\cr
~~~2.28~~~&&~~~1.976305~~~&&~~~-.090631~~~&&~~~.223386~~~&&~~~-.178070~~~\cr
~~~2.29~~~&&~~~1.970443~~~&&~~~-.087694~~~&&~~~.210139~~~&&~~~-.166292~~~\cr
~~~2.30~~~&&~~~1.962762~~~&&~~~-.084833~~~&&~~~.204750~~~&&~~~-.162334~~~\cr
~~~2.31~~~&&~~~1.953283~~~&&~~~-.082321~~~&&~~~.201087~~~&&~~~-.159927~~~\cr
~~~2.32~~~&&~~~1.942236~~~&&~~~-.080266~~~&&~~~.198051~~~&&~~~-.157918~~~\cr
~~~2.33~~~&&~~~1.929886~~~&&~~~-.078616~~~&&~~~.195733~~~&&~~~-.156425~~~\cr
~~~2.34~~~&&~~~1.916495~~~&&~~~-.077331~~~&&~~~.193934~~~&&~~~-.155269~~~\cr
~~~2.35~~~&&~~~1.902327~~~&&~~~-.076385~~~&&~~~.192421~~~&&~~~-.154228~~~\cr
~~~2.36~~~&&~~~1.887647~~~&&~~~-.075764~~~&&~~~.191102~~~&&~~~-.153220~~~\cr
~~~2.37~~~&&~~~1.872718~~~&&~~~-.075460~~~&&~~~.189937~~~&&~~~-.152207~~~\cr
~~~2.38~~~&&~~~1.857789~~~&&~~~-.075402~~~&&~~~.188928~~~&&~~~-.151227~~~\cr
~~~2.39~~~&&~~~1.842973~~~&&~~~-.075403~~~&&~~~.188076~~~&&~~~-.150375~~~\cr
~~~2.40~~~&&~~~1.828310~~~&&~~~-.075456~~~&&~~~.187333~~~&&~~~-.149604~~~\cr
~~~2.41~~~&&~~~1.813840~~~&&~~~-.075563~~~&&~~~.186643~~~&&~~~-.148862~~~\cr
~~~2.42~~~&&~~~1.799604~~~&&~~~-.075726~~~&&~~~.185993~~~&&~~~-.148130~~~\cr
~~~2.43~~~&&~~~1.785641~~~&&~~~-.075944~~~&&~~~.185374~~~&&~~~-.147402~~~\cr
~~~2.44~~~&&~~~1.771977~~~&&~~~-.076183~~~&&~~~.184775~~~&&~~~-.146683~~~\cr
~~~2.45~~~&&~~~1.758616~~~&&~~~-.076435~~~&&~~~.184202~~~&&~~~-.145985~~~\cr
~~~2.50~~~&&~~~1.696441~~~&&~~~-.077883~~~&&~~~.182826~~~&&~~~-.143884~~~\cr
~~~2.55~~~&&~~~1.642101~~~&&~~~-.079563~~~&&~~~.182356~~~&&~~~-.142575~~~\cr
~~~2.60~~~&&~~~1.595003~~~&&~~~-.081199~~~&&~~~.182339~~~&&~~~-.141739~~~\cr
~~~2.65~~~&&~~~1.554151~~~&&~~~-.082690~~~&&~~~.182157~~~&&~~~-.140811~~~\cr
~~~2.70~~~&&~~~1.518180~~~&&~~~-.083756~~~&&~~~.181851~~~&&~~~-.139974~~~\cr
~~~2.75~~~&&~~~1.485366~~~&&~~~-.084280~~~&&~~~.181587~~~&&~~~-.139447~~~\cr
~~~2.80~~~&&~~~1.454568~~~&&~~~-.084734~~~&&~~~.181364~~~&&~~~-.138997~~~\cr
~~~~\infty~~~~~&&~~~1.000000~~~&&~~~-.092878~~~&&~~~.114025~~~&&~~~-.067586~~~\cr
\noalign{\hrule}}
}$$
\vskip 5pt
\centerline {\bf \table{nonpert}}
\vskip 5pt
\vbox{\baselineskip=10pt
\noindent
\centerline{The correction factor $\lambda$ and
non-perturbative coupling derivatives.}
}\baselineskip=15pt
\medskip
\noindent
The interpolation was checked further by a comparison of the
coupling constant shift $\Delta\beta$ to MCRG measurements with
scale two blocking transformations \refs{Hell}{Deck}. In Fig. 3 the data
and our $\Delta\beta$ are shown together with the two-loop prediction.
At high $\beta$-values the measurements from the small $(8^4)$
lattice blocking are slightly inconsistent with those on a large
$(32^4)$ lattice. Our result prefers the large lattice data and is,
in contrast to the two-loop function, in full agreement with the rest
of the data. In addition we have included in the figure data derived
from the critical couplings $g^2_c$ of the deconfinement transition
for \nt-values, which differ by a factor two. They are of course
coinciding with our curve.


 \bigskip
 \par\noindent
{\bf 3. The non-perturbative coupling derivatives}
 \bigskip

\par
The non-perturbative $\beta$-function, which we have just derived,
yields one of the two equations for the determination of the
non-perturbative coupling derivatives. Another one is found
by expressing, via \eqand{ep}{delta}, the pressure in terms of the
derivatives
$$\eqalign{
 { P \over T^4} =
 N\nt^4\biggl\{ &
             \biggl[2g^{-2}-
\biggl({{\partial}\gs^{-2} \over{\partial}\xi}
- {{\partial} \gt^{-2} \over {\partial} \xi}\biggr)\biggr]
  (P_{\sigma}-P_{\tau})       \cr
    &-3 \biggl({{\partial}\gs^{-2} \over{\partial}\xi}
+ {{\partial} \gt^{-2} \over {\partial} \xi}\biggr)
  \biggl[ 2P_0-(P_{\sigma}+P_{\tau}) \biggr]
       \biggr\}~.             \cr}
\EQNO{Press}
$$
In ref.\ref{pressure} it was demonstrated how the pressure may
be calculated as well non-perturbatively without derivatives,
at least on large lattices. To this end one has to calculate
the normalized free energy density from
$${
{f\over T^4}\Big\vert_{\beta_0}^{\beta} =~-3\nt^4\int_{\beta_0}^{\beta}
 {\rm d}\beta' \biggl[ 2P_0-(P_{\sigma}+P_{\tau}) \biggr]~,}
\EQNO{freeenergy}
$$
where $\beta = 4/g^2$, and to use the identity
$$
P=-f~,
\EQNO{identity}
$$
which is valid for homogeneous systems. It was checked for $SU(2)$
\ref{pressure} that
a lattice of size $18^3\times 4$ ( where we have many data
points ) is large enough to justify this approximation.
We have taken advantage of this fact and first calculated the pressure
non-perturbatively on the $18^3\times 4$ lattice by interpolating
and integrating over $[2P_0-(P_{\sigma}+P_{\tau})]$.
Subsequently we used \eqand{sum}{Press} to obtain
$$
{{\partial}\gs^{-2} \over{\partial}\xi} =
g^{-2} - {1 \over 4}{ dg^{-2} \over d\ln a}
+ {1 \over (P_{\sigma}-P_{\tau})} \biggl( - {P \over 2N\nt^4T^4}
+ {3 \over 4}{ dg^{-2} \over d\ln a}[2P_0-(P_{\sigma}+P_{\tau})]
 \biggr)~,
\EQNO{csp}
$$
and
$$
{{\partial}\gt^{-2} \over{\partial}\xi} =
- {{\partial}\gs^{-2} \over{\partial}\xi}
- {1 \over 2}{ dg^{-2} \over d\ln a}~,
\EQNO{cst}
$$
separately. It is clear, that below the critical point
$(4/g^2 < 2.30)$ , where both the pressure $P$ and the plaquette
difference $P_{\sigma}-P_{\tau}$ are small, the error on
${\partial}\gs^{-2} /{\partial}\xi$ becomes relatively large. We
have therefore smoothly interpolated the values in that region.
The resulting non-perturbative derivatives have been listed in
\table{nonpert}. We observe strong deviations from
 the asymptotic predictions \ref{csct} up to large values
 of the coupling $(4/g^2 \simeq 3.0)$.
 \bigskip
 \par\noindent
{\bf 4. Finite size scaling behaviour}
 \bigskip
 \par
{ A. The phase transition region}
 \bigskip
\par
On a finite lattice the correlation length is limited by the
characteristic length scale of the system, \ie~normally the
spatial extension \ns. This limitation prevents the divergence
of the correlation length at the critical point of a second order
deconfinement transition. As a consequence the thermodynamic
observables exhibit close to the transition finite size scaling
properties, which are determined by the critical exponents and
the length scale. According to the universality hypothesis
of Svetitsky and Yaffe \ref{Svet} the exponents of $SU(2)$ should
be the same as those of the three-dimensional Ising model.
\par
Usually finite size scaling analyses are carried out at fixed \nt~,
but varying \ns, see \eg~ref.\ref{Engc}. The spatial extension
determines then obviously the scale. In investigations, where results
from lattices with varying \ns~and \nt~are compared, the dimensionless
ratio
$$
{{\ns} \over{\nt}} = LT~,
\EQNO{ratio}
$$
where $L = \ns a$, has turned out to be the appropriate
variable. This was observed already in studies of the heavy quark
potential \ref{Enge}, and later on confirmed by the behaviour of the
Binder cumulant \ref{asymp}.
\par
We may check this idea further by looking at the interaction measure
$\Delta$, \eq{delta}, for various \ns~ and \nt~ in the neighbourhood
of the critical point, as a function of $T/T_c$. Note, that here
the non-perturbative $\beta$-function enters in both the quantity
$\Delta$ and the scale $T/T_c$, but no single coupling derivative.
In Fig. 4 we see a clear coincidence of $\Delta$-data belonging
to the same ratio $\ns/\nt$.
\par
The finite size dependence of the energy density is related to the
singular part $f_s$ of the free energy density. Since the critical
exponent $\alpha$ of the specific heat is relatively small
$(\alpha \approx 0.11)$ \ref{Engc}, one expects that the regular
parts in the specific heat and in the energy density are dominating.
Correspondingly, we make the following ansatz for the energy density
close to $T_c$
$$
{\epsilon \over T^4} = \biggl({\epsilon \over T^4}\biggr)_{reg} +
\biggl({{\ns} \over{\nt}}\biggr)^{(\alpha-1)/\nu}Q_{\epsilon}
\biggl( t \biggl({{\ns} \over{\nt}}\biggr)^{1/\nu} \biggr)~,
\EQNO{entc}
$$
where we have already neglected irrelevant scaling fields. Here
$$
t = (T-T_c)/T_c
\EQNO{rtemp}
$$
is the reduced temperature, $Q_{\epsilon}$ the scaling function
of the energy density
and $\nu$ the critical exponent of the correlation length.
 \bigskip
 \par
{ B. The weak coupling region}
 \bigskip
\par
For small bare couplings $g^2$, in the vicinity of the continuum
limit, the energy density may be approximated by its weak coupling
expansion. To obtain it up to order $g^2$, we first derive
$\epsilon /T^4$ from \eqand{ep}{delta}
$$\eqalign{
 { \epsilon \over T^4} =
 3N\nt^4\biggl\{ &
             \biggl[2g^{-2}-
\biggl({{\partial}\gs^{-2} \over{\partial}\xi}
- {{\partial} \gt^{-2} \over {\partial} \xi}\biggr)\biggr]
  (P_{\sigma}-P_{\tau})       \cr
    &+ \biggl({{\partial}\gs^{-2} \over{\partial}\xi}
+ {{\partial} \gt^{-2} \over {\partial} \xi}\biggr)
  \biggl[ 2P_0-(P_{\sigma}+P_{\tau}) \biggr]
       \biggr\}~.             \cr}
\EQNO{Energ}
$$
Into this equation we insert the weak coupling expansions of the
plaquettes for $SU(N)$ lattice gauge theory \ref{weak}
$$
P = g^2{{N^2-1} \over {N}}P^{(2)} + g^4(N^2-1)P^{(4a)}
   + g^4{{(2N^2-3)(N^2-1)} \over {N^2}}P^{(4b)} + O(g^6) ,
\EQNO{plaq}
$$
and the asymptotic values of the coupling
derivatives \ref{csct}
$$\eqalign{
{{\partial}\gs^{-2} \over{\partial}\xi}
- {{\partial}\gt^{-2} \over{\partial}\xi}~= &~
  {N^2-1 \over N} \cdot 0.146711 -N \cdot 0.019228~,   \cr
{{\partial}\gs^{-2} \over{\partial}\xi}
+ {{\partial}\gt^{-2} \over{\partial}\xi}~= &~b_0 =
 {{11N}\over {48\pi^2}} ~.                             \cr}
\EQNO{deriva}
$$
The coefficients of the perturbative formula
$$
 { \epsilon \over T^4} =
 a_0 + a_1 g^2 + O(g^4)~,
\EQNO{peren}
$$
are related to those of the plaquettes in the following way
$$\eqalign{
 a_0 =~
 6(N^2-1) \nt^4 & \Delta P^{(2)}_{\sigma \tau}~,   \cr
 a_1 =~
 6(N^2-1) \nt^4 &\biggl\{
 N\Delta P^{(4a)}_{\sigma \tau} + { 2N^2-3 \over N}
 \Delta P^{(4b)}_{\sigma \tau} - 0.146711 { N^2-1 \over 2N }
 \Delta P^{(2)}_{\sigma \tau}                          \cr
       & + 4N~[~0.000499 \Delta P^{(2)}_{0\sigma}
             +0.005306  \Delta P^{(2)}_{0\tau}  ]
       \biggr\}~,             \cr}
\EQNO{encoe}
$$
where
$$
\Delta P_{\alpha \beta} = P_{\alpha} - P_{\beta}~,
\EQNO{delp}
$$
was used as an abbreviation and $P_0$ is the plaquette on the
respective symmetric $\ns^4$ lattice. The coefficients $a_0, a_1$
approach with increasing \ns~and \nt~the corresponding continuum
values
$$\eqalign{
a_0 = &~{\pi^2 \over 15} (N^2-1)~~,        \cr
a_1 = &-{1 \over 48} (N^2-1)N ~.        \cr}
\EQNO{acon}
$$
\par
We have calculated both coefficients on a large number of $\ns^3\times
\nt$ lattices. In Figs. 5 and 6 they are plotted for $SU(2)$ at
fixed ratios \ns/\nt~as a function of \nt. We observe, that $a_0$
has essentially
the same \nt-dependence for all ratios \ns/\nt~and that the
thermodynamic limit $\ns \rightarrow \infty$ is almost reached already
for $\ns/\nt > 4$. For a lattice calculation with an energy density
as close as possible to the continuum limit one should therefore
choose \nt~as large as possible at a moderate ratio \ns/\nt.
\par
Asymptotically, for $\ns \rightarrow \infty$ and large \nt~
the leading terms of $a_0$ are given for $SU(N)$ by
$$
a_0 = (N^2-1) \biggl [{\pi^2 \over 15}
 + {2 \pi^4 \over 63} \cdot {1 \over \nt^2} +
 O\biggl ({1 \over \nt^4} \biggr ) \biggr ]~.
\EQNO{anull}
$$
The \ns- and \nt-dependences of $a_1$
are more complicated.
For $\ns \rightarrow \infty$ and large \nt~
the second coefficient approaches the continuum value from
below, in contrast to $a_0$.
For fixed \ns/\nt, both $a_0$ and $a_1$ cross their respective
continuum limit lines, but at different \nt-values.
 \vfill\eject
 \bigskip
 \par\noindent
{\bf 5. Numerical results for the energy density}
 \bigskip
\par
With our results for the non-perturbative derivatives and the
$\beta$-function we are now able to reevaluate the energy density
with the plaquette values from refs. \refs{Engc}{Engd}.
In Fig. 7 we show $\epsilon/T^4$ in the vicinity of the deconfinement
transition as a function of $T/T_c$. Like in the case of the
interaction measure $\Delta$ we observe that at fixed temperature
$\epsilon/T^4$ depends only on the ratio $\ns/\nt$.
From the ansatz, \eq{entc}, the following scaling behaviour
is then expected at $t = 0$
$$
{\epsilon \over T^4} = \biggl ({\epsilon \over T^4} \biggr )_{\infty}
+ \biggl ({\ns \over \nt}\biggr)^{(\alpha-1)/\nu} \cdot Q_{\epsilon}(0)
~.
\EQNO{entcr}
$$
 We have made fits of this form to data at $T_c$ coming from seven
 different lattice sizes.
 Fixing the critical exponents to those of the $3d$ Ising
 model, \ie
$$
(\alpha -1) /\nu = -1.41~,
\EQNO{crie}
$$
we find for the value of the energy density in the limit
$\ns/\nt \rightarrow \infty$ ( the regular term at $T_c$ )
$$
\biggl ({\epsilon \over T^4}\biggr )_{\infty} = 0.256 (23)~.
\EQNO{cont}
$$
%
%
Our result is somewhat lower than the one found previously in ref.
\ref{Mitr}, where perturbative derivatives were used.
A plot of the fit and the data are shown in Fig. 8.
If we consider the exponent of $(\ns/\nt)$ in \eq{entcr} as a free
parameter we find as best value
$$
(\alpha -1) /\nu = -1.58~.
\EQNO{ourv}
$$
This entails a slight change in the continuum limit value, \eq{cont},
which however was already taken into account in the error estimate.
\par
In the confined, low temperature phase an often used model for the
energy density is that of a free gas of massive glueballs of various
spins and parities. Assuming Boltzmann statistics the energy density
of such a glueball gas is given by
\medskip
$$
{\epsilon_{gb} (T) \over T^4} = {1\over {2\pi^2}}\sum_i
 d_i \biggl({{M_i}\over T} \biggr)^3 \biggl [K_1
\biggl({{{M_i}\over T}} \biggr )+{{3T}\over
 {M_i}}K_2 \biggl({{M_i}\over T}\biggr) \biggr ]~~.
\EQNO{glueba}
$$
\medskip
In pure $SU(2)$ lattice gauge theory the following glueball states
have been identified :
 $M_{O^+}/T_c\sim 5.5$,
 $M_{2^+}/T_c\sim 8.0$,
 $M_{O^-}/T_c\sim 9.5$,
 and
 $M_{2^-}/T_c\sim 10.2$ \ref{glueb}.
 At the critical temperature the energy
 density of an ideal gas of these glueballs is then found to be
$$
\epsilon_{gb} (T_c) / T_c^4 = 0.050~,
\EQNO{glbc}
$$
\ie~a factor five less than our estimate, \eq{cont}. The presence
of strong glueball interactions is therefore to be expected in the
confinement phase.
\par
Comparing the critical energy density with that of an asymptotically
free gluon gas
$$
\epsilon_{g} (T) / T^4 = \pi^2/5~,
\EQNO{stbo}
$$
we find a difference of nearly a factor eight. This means, that
also in the deconfinement phase close to the transition strong
interactions of the constituents of the gluon plasma are anticipated.
\par

$$\vbox{\offinterlineskip
\halign{
\strut\vrule     \hfil $#$ \hfil  &
      \vrule # & \hfil $#$ \hfil  &
      \vrule # & \hfil $#$ \hfil  &
      \vrule # & \hfil $#$ \hfil  &
      \vrule # & \hfil $#$ \hfil
      \vrule \cr
\noalign{\hrule}
  ~4/g^2~
&&~N_{meas}~
&&~P_{\sigma}~
&&~P_{\tau}~
&&~<\mid L \mid >~\cr
\noalign{\hrule}
%
%
%
~~~2.55~~~&&~~~24840~~~&&~~~.3384811(34)~~~&&~~~.3384526(34)~~~&&~~~.07344(19)~~~\cr
~~~2.60~~~&&~~~17750~~~&&~~~.3298618(39)~~~&&~~~.3298233(37)~~~&&~~~.09524(11)~~~\cr
~~~2.65~~~&&~~~42304~~~&&~~~.3218638(23)~~~&&~~~.3218192(22)~~~&&~~~.11147(06)~~~\cr
~~~2.74~~~&&~~~91900~~~&&~~~.3086725(15)~~~&&~~~.3086204(14)~~~&&~~~.13600(04)~~~\cr
\noalign{\hrule}}
}$$
\vskip 5pt
\centerline {\bf \table{32}}
\vskip 5pt
\vbox{\baselineskip=10pt
\noindent
\centerline{Results from Monte Carlo calculations
on a $32^3 \times 8$ lattice.}
}\baselineskip=15pt
 \bigskip
 \par\noindent
%
Let us now consider the high temperature phase. In Fig. 9a we show
the approach of the energy density to the leading weak coupling term
$a_0 = \epsilon^{wc}/T^4$ - the lattice Stefan-Boltzmann limit - for two
lattices. The $32^3 \times 8$ data are new. Here the lattice
size was chosen such as to have a large \nt-value and a reasonable
$\ns/\nt$-ratio at the same time. Details of these data are presented
in \table{32}. We see from Fig. 9a that the data even at
$T/T_c \approx 2$
are still about $30\%$ below that of a gas of non-interacting
$SU(2)$ gluons and that
the deviations from it at fixed temperature
are only slightly \nt-dependent. Yet, we are still not in the weak
coupling region, since the differences cannot be explained by the
next-to-leading order weak coupling term. This does not come as a
surprise, because in the $T$-range shown, the coupling derivatives
are still non-perturbative.
\par
In Fig. 9b we have plotted the difference between $\epsilon^{wc}/T^4$
and the energy density data, showing again that the lattice
Stefan-Boltzmann limit is clearly approached from below.
 A comparison of $(\epsilon^{wc}-
\epsilon)/T^4$ to finite temperature perturbation theory with only
one term proportional to the temperature dependent running coupling
constant leads to a large value for $g^2(T)$ of the order of $4 \sim 5$.
It seems therefore that at presently accessible temperature values
low order perturbation theory is not applicable.
 \bigskip
 \par\noindent
{\bf 6. Conclusions and discussion }
 \bigskip
\par
We have studied the finite size scaling behaviour of the energy
density in $SU(2)$ gauge theory at finite temperature. We find
that the energy density at the critical point is only about $1/8$
of the ideal gas value and also at $T \simeq 2T_c$
it is still $30\%$ below this limit. The discrepancy between this
result and earlier findings of a rapid approach to the ideal gas
limit can be traced
back to our non-perturbative definition of the pressure,
\eq{freeenergy}. The additional use of a non-perturbative
$\beta-$function in the definition of $(\epsilon-3P)/T^4$ is a minor
modification to this. In fact the non-perturbative calculation of
$P/T^4$ through \eq{freeenergy} yields above $T_c$ a smaller value
than \eq{Press} with perturbative coupling derivatives. This is
reflected in a strong deviation of our results for $\partial
g_{\sigma,\tau}^{-2}/\partial\xi$ from the perturbative values
( see \table{nonpert}). A similar result was obtained in $SU(3)$,
where these derivatives were directly calculated from Wilson loops
on anisotropic lattices \ref{Burg}.
\par
Some uncertainty on the absolute value of the pressure and
consequently also the energy density is due to our approximation
$P/T^4 \equiv 0$ for $T \simeq 0.8T_c$. This is suggested by
the Monte Carlo data and seems also justified because the confinement
phase consists only of heavy glueball states. We also note that
unlike in the perturbative definition of thermodynamic quantities
the non-perturbative approach does not automatically ensure that
the ideal gas limit is reached for $4/g^2 \rightarrow \infty$. However,
it was checked in \ref{pressure}, that the order $g^2$ correction
to \eq{freeenergy}
in the infinite temperature limit has the correct perturbative form.
\par
In the near future more Monte Carlo calculations in the high
temperature region are needed to clarify the nature of
the approach to the region where
finite temperature perturbation theory is valid.
\bigskip

 \vfill\eject
\par\noindent
{\bf  Acknowledgements:}
We are indebted to the HLRZ J\"ulich
for providing the
necessary computer time. We thank J.~Rank for the
data from his simulations on the $32^3 \times 8$ lattice.
K.~R. acknowledges support by the Stabsabteilung Internationale
Beziehungen, KFA Karlsruhe and Komitet Bada\'n Navkowych (KBN).
One of us (K.~R.) is indebted to H.~Satz for interesting
discussions and suggestions.
\vskip 15pt
 \bigskip
{\centerline{\bf References}}
\vskip 15pt
\item{\reftag{pressure})}
J.~Engels, J.~Fingberg, F.~Karsch, D.~Miller and M.~Weber, \PL B252
(1990) 625.
\item{\reftag{asymp})}
J.~Fingberg, U.~Heller and F.~Karsch, \NP B392 (1993) 493.
\item{\reftag{Engc})}
J.~Engels, J.~Fingberg and M.~Weber, \NP B332 (1990) 737.
\item{\reftag{Engd})}
J.~Engels, J.~Fingberg and D.~E.~Miller, \NP B387 (1992) 501.
\item{\reftag{Akemi})}
K.~Akemi at al. \PRL 71 (1993) 3063.
\item{\reftag{Kh})}
A.~X.~El-Khadra, G.~Hockney, A.~S.~Kronfeld and P.~B.~
Mackenzie, \PRL 69, (1992) 729;
\item{\reftag{Hell})}
U.~Heller and F.~Karsch, \PRL 54 (1985) 1765. 
\item{\reftag{Deck})}
K.~M.~Decker and P.~de Forcrand, \NP B(Proc. Suppl.) 17 (1990) 567.
\item{\reftag{csct})}
F.~Karsch, \NP B205 (1982) 285. 
\item{\reftag{Svet})}
B.~Svetitsky and G.~Yaffe, \NP B210 [FS6] (1982) 423.
\item{\reftag{Enge})}
J.~Engels, F.~Karsch and H.~Satz, \NP B315 (1989) 419.
\item{\reftag{weak})}
U.~Heller and F.~Karsch, \NP B251 [FS13] (1985) 254. 
\item{\reftag{Mitr})}
J.~Engels, J.~Fingberg and V.~Mitrjushkin, \PL B298 (1993) 154.
\item{\reftag{glueb})}
C.~Michael and M.~Teper, Phys. Lett.B199 (1987) 95.
\item{\reftag{Burg})}
G.~Burgers, F.~Karsch, A.~Nakamura and I.~O.~Stamatescu, \NP B304 (1988) 587.

 \vfill\eject
 \bigskip
 \par\noindent
{\bf Figure Captions:}
\medskip
\item{\bf 1.}{The measured critical temperature $1/(\nt R)$ for
$\lambda = 1$ ( circles) and $T_c/\Lambda_L$ \break
( squares) obtained after
correction with the factor $\lambda$ ,vs. $4/g^2$ for
different \nt. The solid line results from our interpolation
of $\lambda$, the dashed-dotted line from the asymptotic form
$\lambda_{as}$. The dashed line is the fit result from \eq{fit2}.
}
\item{\bf 2.}{The sum of derivatives
$({\partial}\gs^{-2}/{\partial}\xi
+ {\partial}\gt^{-2}/{\partial}\xi)$ at $\xi=1$.
The solid line is our \break non-perturbative result,
the dashed-dotted line comes from $\lambda_{as}$ and the dashed line
from the two-loop approximation.
}
\item{\bf 3.}{The coupling constant shift $\Delta\beta$ from scale two
blocking transformations vs. $4/g^2$, plotted at the larger of the two
values. The data were measured by using two different lattices
(diamonds, triangle) \ref{Hell}, blocking on one large lattice
(crosses) \ref{Deck} or obtained from $T_c$ (circles) \ref{asymp}.
The notation of the lines is the same as in Fig. 2.
}

\item{\bf 4.}{The interaction measure
$\Delta = (\epsilon-3P)/T^4$ for
various \ns~ and \nt~ in the neighbourhood
of the critical point, as a function of $T/T_c$. The necessary
plaquette data were taken from refs.\refs{Engc}{Engd}.
}

\item{\bf 5.}{The first coefficient $a_0$ of the weak coupling
expansion of $\epsilon/T^4$ versus $\nt^{-2}$
for the fixed ratios $\ns/\nt=2,4,8,16$ (long dashes,
dotted-dashed,dotted and solid lines). The continuum value is shown
as a line of short dashes; the leading term in $\nt^{-2}$ as a
solid straight line. All numbers are for $SU(2)$.
}

\item{\bf 6.}{The second coefficient $a_1$ of the weak coupling
expansion of $\epsilon/T^4$ versus $\nt$
for the fixed ratios $\ns/\nt=2,4,6,8$ (long dashes,
dotted-dashed,solid and dotted lines). The continuum value is shown
as a line of short dashes. All numbers are for $SU(2)$.
}

\item{\bf 7.}{The energy density $\epsilon /T^4$
for various \ns~ and \nt~ in the neighbourhood
of the critical point, as a function of $T/T_c$.
}

\item{\bf 8.}{The energy density at $T_c$ as a function of
$(\ns/\nt)^{(\alpha-1)/\nu}$ with the critical exponents from
the $3d$ Ising model. The straight line is a fit with the
scaling form, \eq{entcr}. The notation for the data is like in
Fig. 7, the diamond is from a $26^3 \times 4$ lattice.
}

\item{\bf 9.}{Comparison of the leading term of
the weak coupling expansion $a_0 = \epsilon^{wc}/T^4$ to
data for $\epsilon/T^4$ on two different lattices,
versus $T/T_c$. Fig. 9a shows the data directly, Fig. 9b the
difference to the respective weak coupling limit.
}

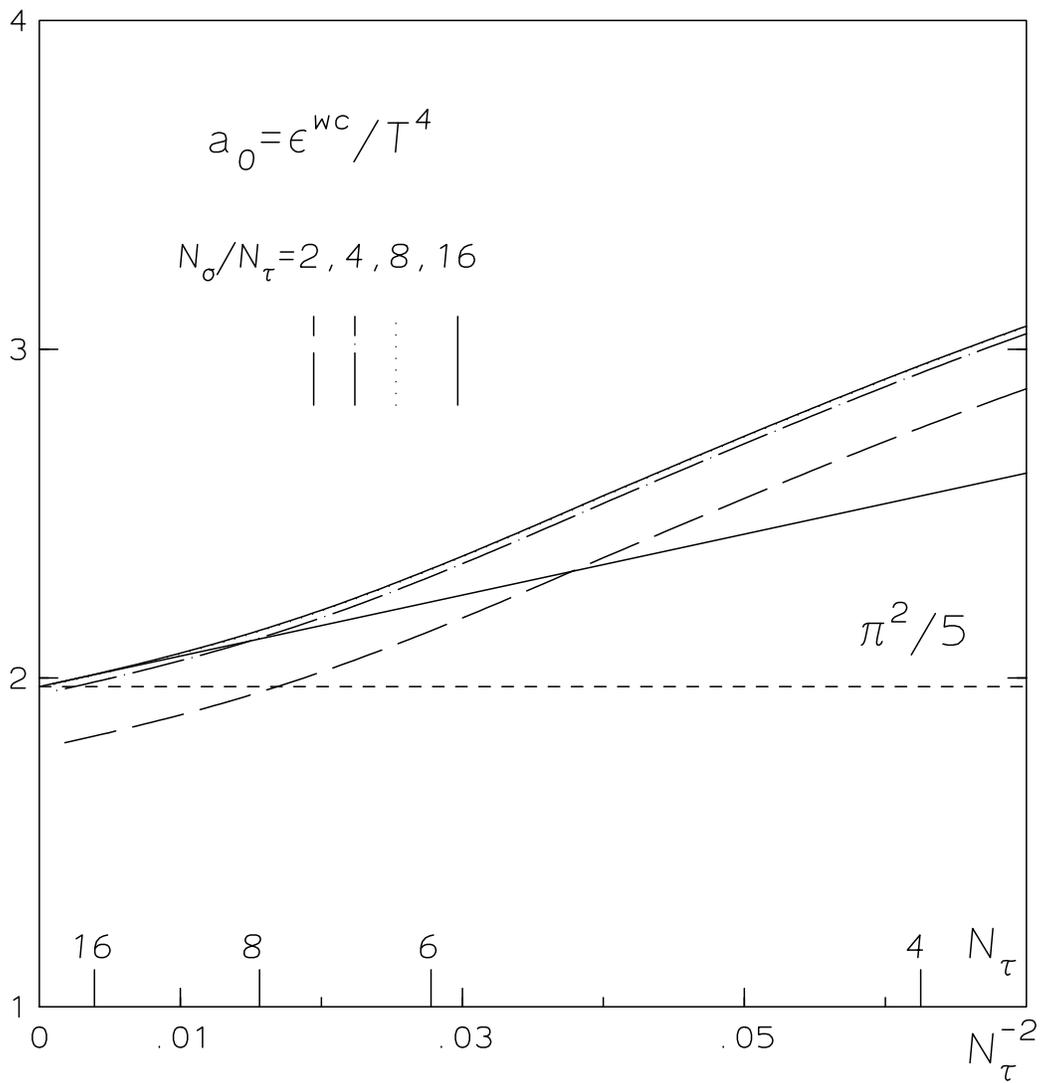
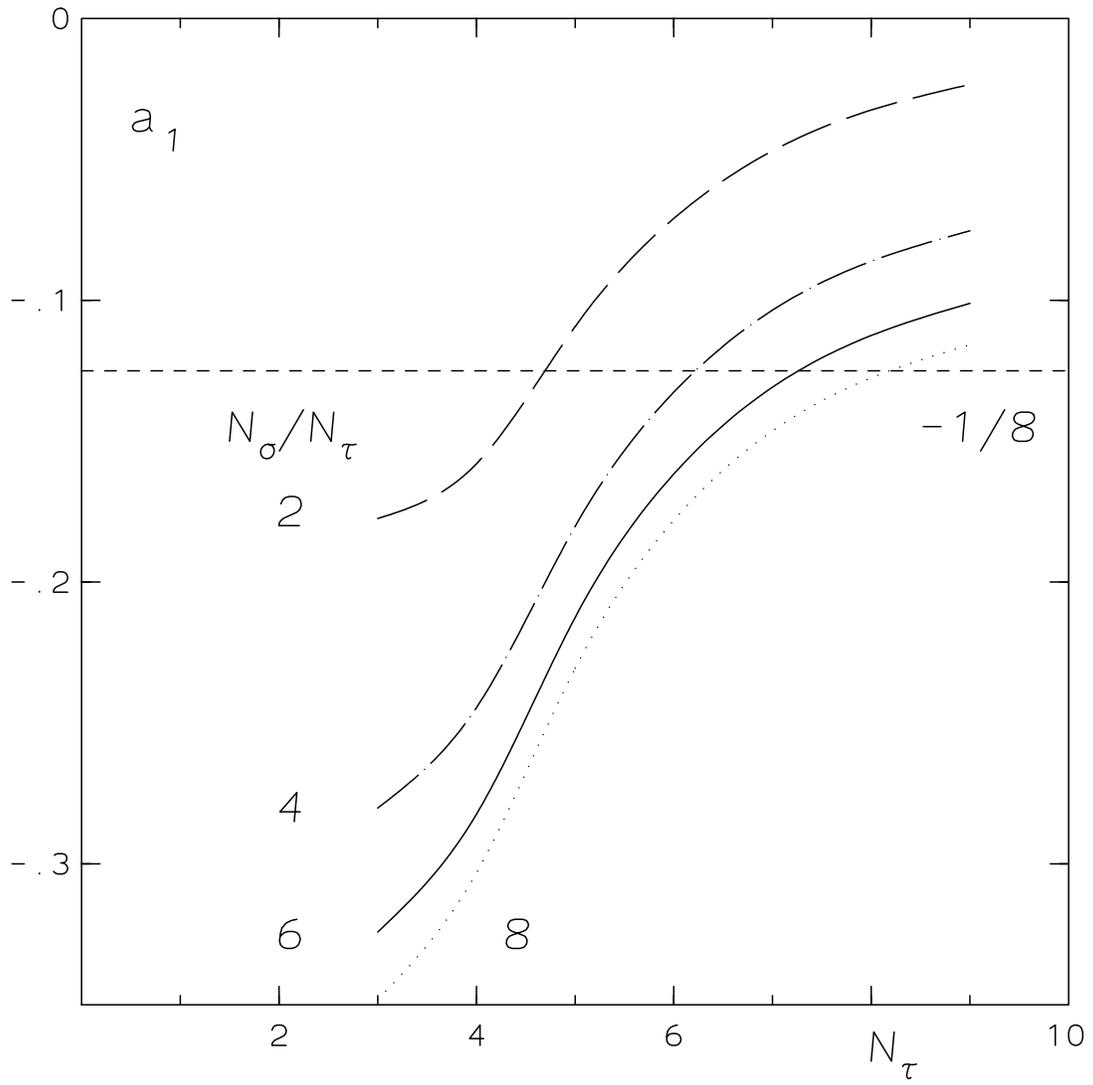
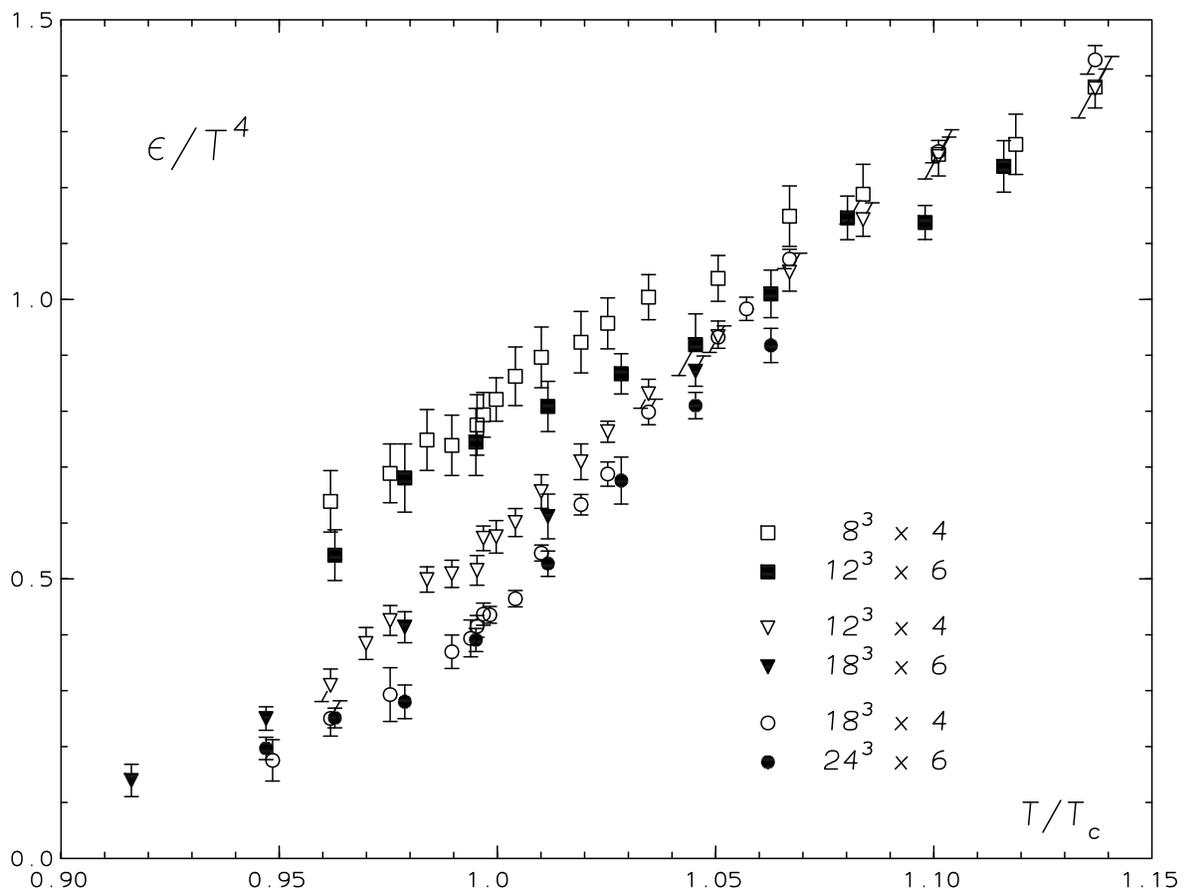
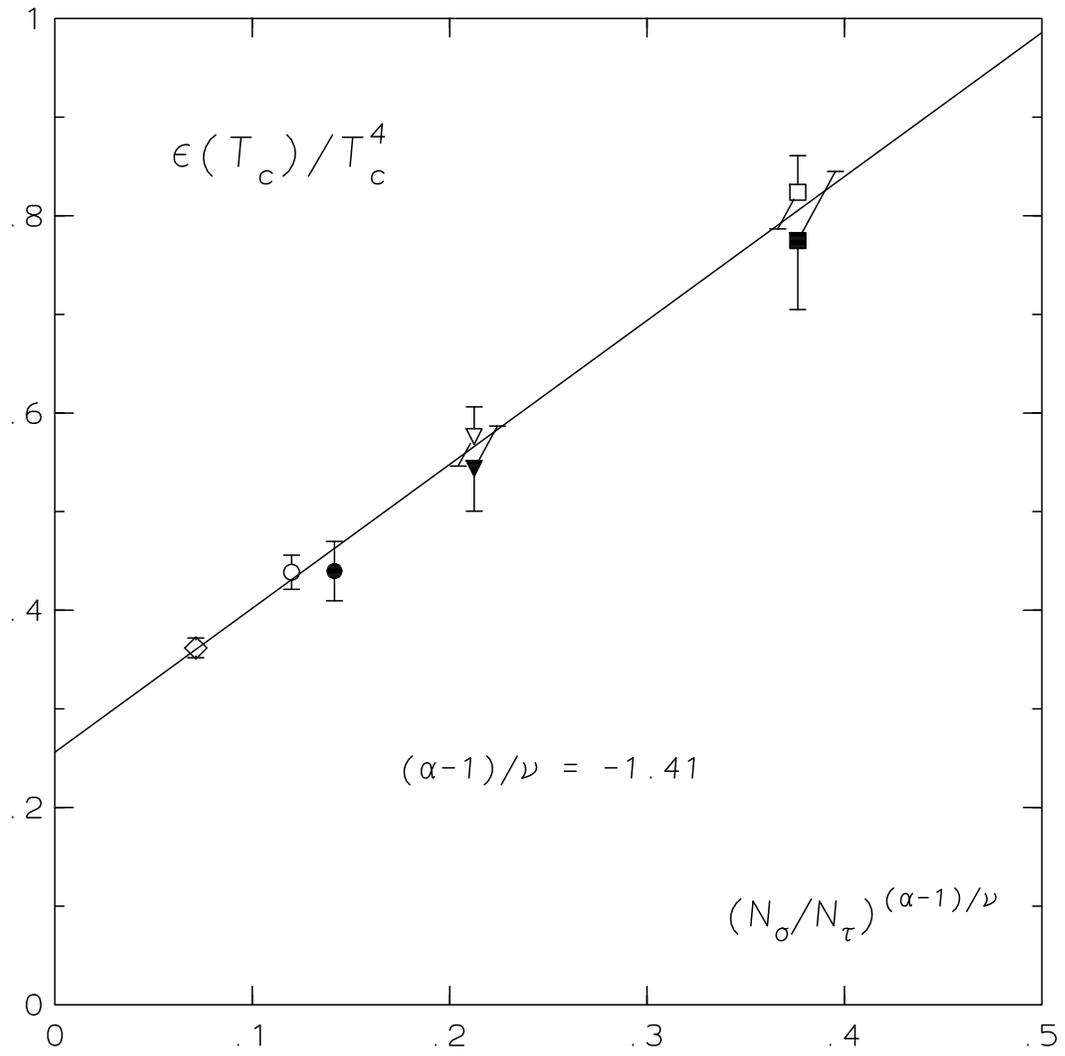
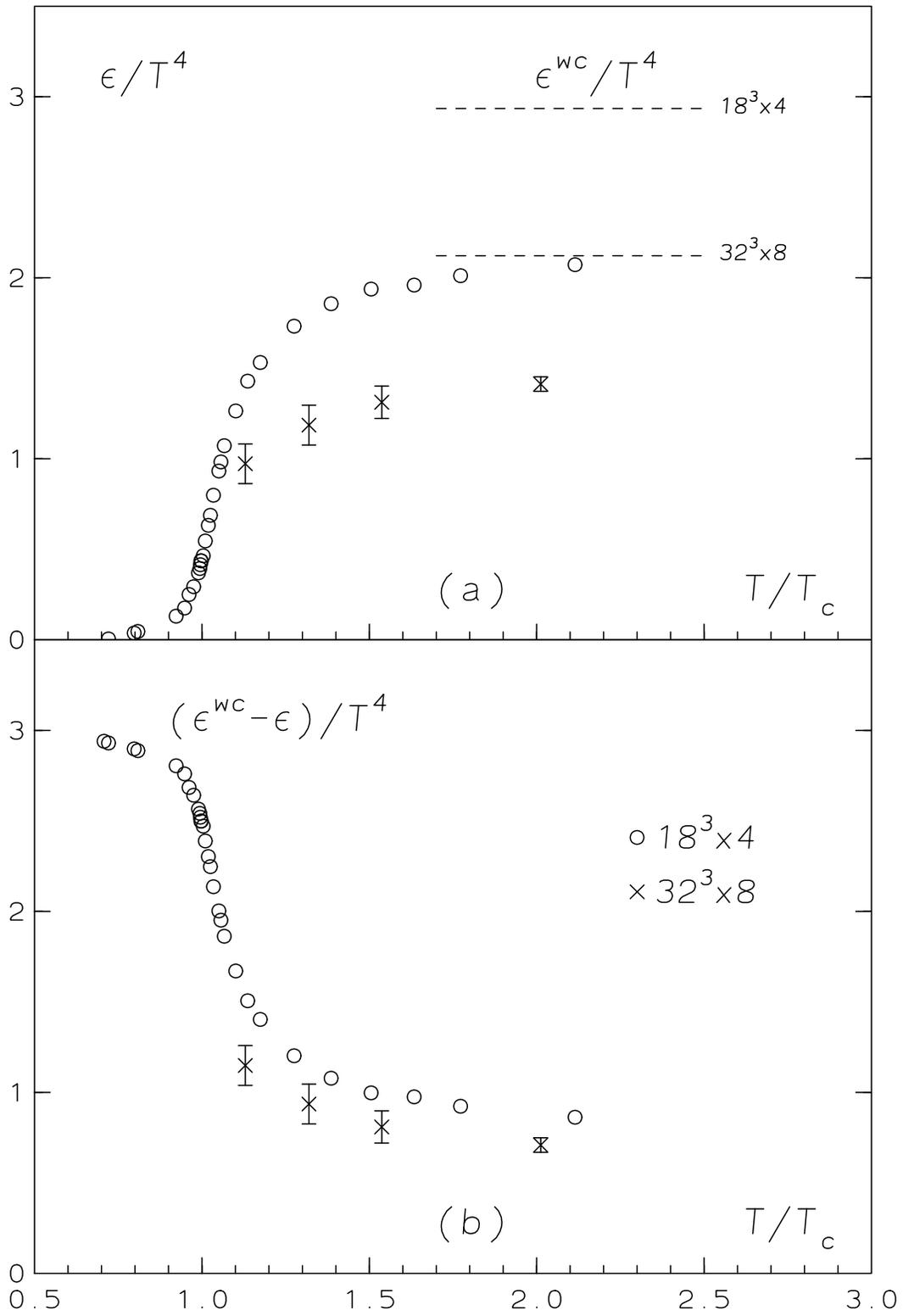
\end